\documentclass[a4paper]{jpconf}
\usepackage{graphicx}

\begin{document}
\title{Variance-Gamma (VG)  model: Fractional Fourier Transform (FRFT)}

\author{A H Nzokem$^1$}

\begin{abstract}
The paper examines the Fractional Fourier Transform (FRFT) based technique as a tool for obtaining the probability density function and its derivatives, and mainly for fitting stochastic model with the fundamental probabilistic relationships of infinite divisibility. The probability density functions are computed, and the distributional proprieties are reviewed for Variance-Gamma (VG) model. The VG model has been increasingly used as an alternative to the Classical Lognormal Model (CLM) in modelling asset prices. The VG model was estimated by the FRFT. The data comes from the SPY ETF historical data. The Kolmogorov-Smirnov (KS) goodness-of-fit shows that the VG model fits the cumulative distribution of the sample data better than the CLM. The best VG model comes from the FRFT estimation.
\end{abstract}

\section{Introduction}
Several empirical studies have shown that asset returns are often characterized by leptokurtosis and asymmetry. These facts provide evidence suggesting the assumptions of the Classical Lognormal Model (CLM) are not consistent with the empirical observations. A natural generalization of the CLM is the method of subordination\cite {clark1973subordinated, hurst1997subordinated}, which has been used to reduce the theoretical-empirical gap. The subordinated process is obtained by substituting the physical time in the CLM by any independent and stationary increments random process, called the subordinator. If we consider the random process to be a Gamma process, we have a Variance Gamma (VG) model, which is the model the paper will be investigating. The Variance Gamma (VG) model was proposed by Madan\cite{madan1990variance}. In contrast to the CLM, the VG model does not have an explicit closed-form of the probability density function and its derivatives. In the paper, the VG model has five parameters: parameters of location ($\mu$), symmetric ($\delta$), volatility ($\sigma$), and the Gamma parameters of shape ($\alpha$) and scale ($\theta$). The VG model density function is proven to be (\ref {eq:l1}).
\begin{equation}
f(y) =\frac {1} {\sigma\Gamma(\alpha) \theta^{\alpha}}\int_{0}^{+\infty} \frac{1}{\sqrt{2\pi v}}e^{-\frac{(y-\mu-\delta v)^2}{2v\sigma^2}}v^{\alpha -1}e^{-\frac{v}{\theta}} \,dv \label{eq:l1}
 \end{equation}
\noindent 
The integral (\ref {eq:l1}) makes it difficult to utilize the density function and its derivatives, and to perform the Maximum likelihood method. However, in the literature, many studies have found a way to circumvent the lack of closed form by decreasing the number of parameters and using approximation function or analytical expression with modified Bessel function. In fact, \cite{madan1987chebyshev} developed a procedure to approximate  (\ref {eq:l1}) by Chebyshev Polynomials expansion. \cite{madan1998variance} and \cite{seneta2004fitting} got (\ref {eq:l1}) by analytical expression with modified Bessel function of second kind and third kind respectively. \cite {Mercuri2010OptionPI} got (\ref {eq:l1}) through Gauss-Laguerre quadrature approximation with Laguerre polynomial of degree $10$.\cite{hurst1997subordinated} used the Fast Fourier Transform (FFT).\\
\noindent 
The Fractional Fourier Transform (FRFT) will be implemented on the Fourier Transform of the VG model function (\ref {eq:l1}) and its derivatives. The paper is structured as follows; the next section presents the analytical framework. The third section presents the Variance Gamma (VG) model and the sample data before performing the parameter estimations of the VG model and the Kolmogorov-Smirnov (KS) goodness-of-fit test.

\section{Analytical Framework}
\subsection{Fast Fourier Transform (FFT) }
\noindent
The continuous Fourier transform (CFT) of function $f(t)$ and its inverse are defined by:
\begin{equation}
 F[f](x) = \int_{-\infty} ^{+\infty}\! f(y)e^{-ixy} \, \mathrm{d}y \label {eq:l2}  \quad  
 f(x) = \frac{1} {2\pi}\int_{-\infty}^{+\infty}\! F[f](y) e^{ixy}\, \mathrm{d}y  
\end{equation}
where i is the imaginary unit.\\
 The Fast Fourier Transform (FFT) is commonly used to evaluate the integrals (\ref {eq:l2}) . The fundamentally inflexible nature \cite {Bailey1994AFM} of FFT is the main weakness of the algorithm. The advantages of computing with the FRFT \cite {Bailey1994AFM} can be found at three levels: (1) both the input function values $f(x_k) $ and the output transform values $F[f](x_k) $ are equally spaced; (2) a large fraction of $f(x_k) $ are either zero or smaller than the computer machine epsilon; and (3) only a limited range of $f(x_k) $ are required. \\
 The FRFT is set up on $n$-long sequence ($x_1$, $x_ {2} $, \dots, $x_{n}$)  and  the Discrete Fourier Transform (DFT), $G_k(x,\delta)$, can be shown in \cite {bailey1991fractional} to be a composition of ${DFT}^ {-1} $ and ${DFT} $.
 \begin{equation}
 G_k(x,\delta)=\sum_{j=0}^{n-1}\! x_ie^{-2\pi i jk\delta} \hspace{5mm}
   \hbox{$0\leq k<M$}  \quad   G_k(x, \delta)=e^{-\pi ik^2\delta}{DFT}_k^{-1}[{{DFT}_j(y){DFT}_j(z)}] 
   \label {eq:l3}
\end{equation} 
\noindent 
where ${DFT}^ {-1} $ is the inverse of the Discrete Fourier Transform (DFT). We assume that $F[f](t)$ is zero outside the interval $[-\frac{a}{2}, \frac{a}{2}]$, and $\beta=\frac{a}{n} $ is the step size of the $n$ input values $F[f](t) $; we define $t_j=(j-\frac{n}{2}) \beta$ for $ 0 \leq j <n$. We have also $\gamma$ as the step size of the $n$ output values of $f(t)$ and $x_k=(k-\frac{n}{2}) \gamma$ for $ 0 \leq k <n$. By choosing the step size $\beta$ on the input side and the step size $\gamma$ in the output side, we fix the FRFT parameter $\delta=\frac{\beta\gamma}{2\pi}$.\\
 \noindent
The density function $f$ at $x_k$ can be written as (\ref{eq:l4}). The proof is provided in \cite{nzokem2021fitting}.
 \begin{equation}
 \hat{f}(x_k) = \frac{\gamma} {2\pi}e^{-\pi i(k-\frac{n}{2})n\delta}G_{k}(F[f](y_{j})e^{-\pi i jn\delta}),-\delta) 
 \label {eq:l4}
 \end{equation}  
\noindent 
 In order to perform $f(t)$ function from the Fourier Transform (FT), we assume $a=20$, $n=2048$, $\beta=\gamma=\frac{a}{n} $. For more detail on FRFT, see  \cite {bailey1991fractional}.

\subsection{Variance Gamma (VG) Distribution}
\begin{figure}[ht]
	 \begin{minipage}[b]{0.32\linewidth}
	    \vspace{-0.2cm}
	   \centering
	   \includegraphics[width=\textwidth]{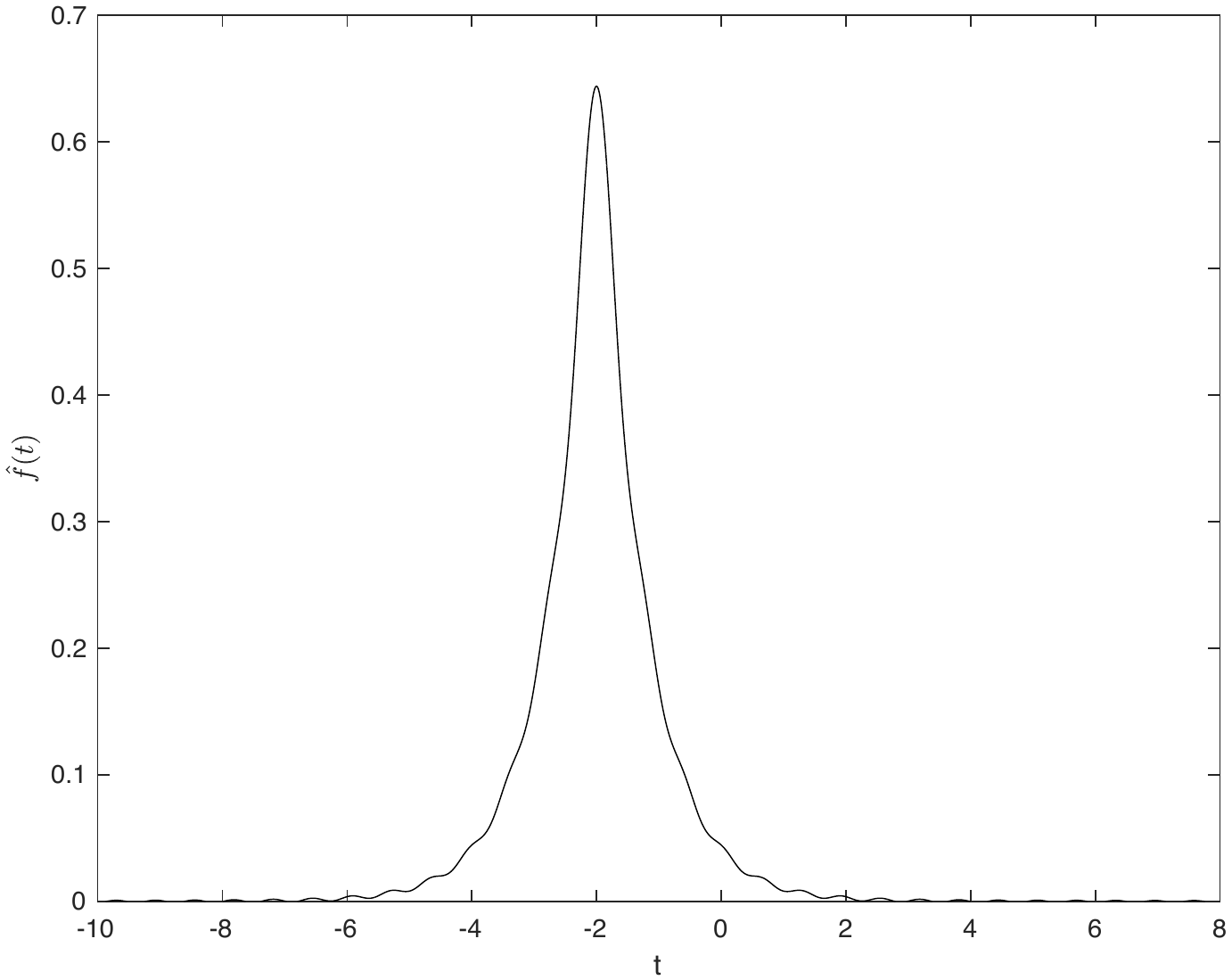}
	    \vspace{-0.7cm}     
    \caption{$\hat{f}$: $\mu=-2$, \\ $\delta=0$, $\sigma=1$, $\alpha=1$, $\theta=1$}
         \label{fig21}
	\end{minipage}
 \hfill 	
	 \begin{minipage}[b]{0.32\linewidth}
	 \vspace{-0.2cm}
	   \centering
	   \includegraphics[width=\textwidth]{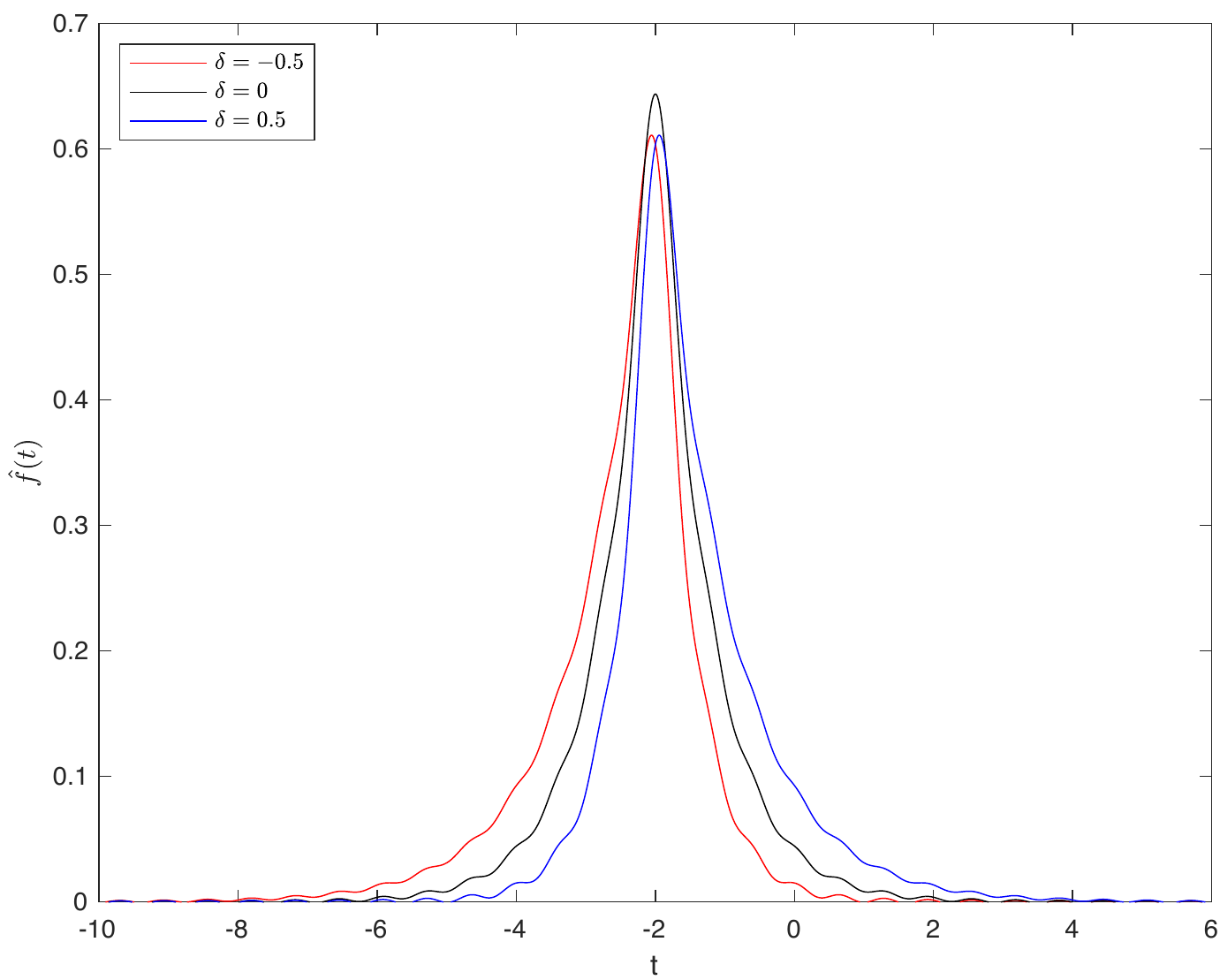}
	    \vspace{-0.7cm}
    \caption{$\hat{f}(t)$ \\and symmetric parameter($\delta$)}
         \label{fig22}
	\end{minipage}
 \hfill	
	 \begin{minipage}[b]{0.32\linewidth}
	 \vspace{-0.2cm}
	   \centering
	   \includegraphics[width=\textwidth]{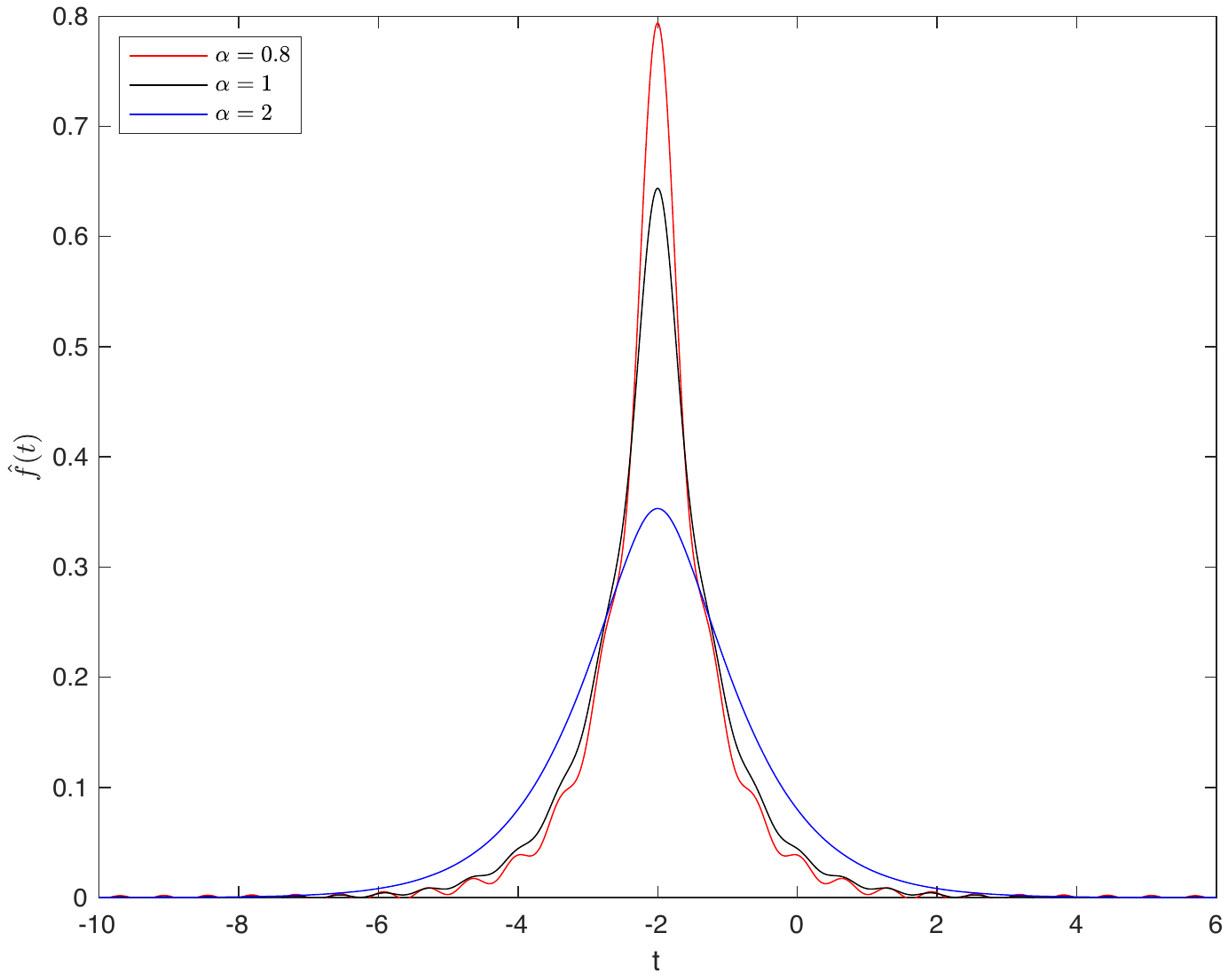}
	   \vspace{-0.7cm}
     \caption{$\hat{f}(t)$ \\and shape parameter($\alpha$)}
         \label{fig23}
	\end{minipage}
\vspace{-0.4cm}
\end{figure}

\begin{equation}
 X =\mu + \delta V +\sigma \sqrt{V}Z  \quad  Z \sim N(0,1) \quad  V \sim \Gamma(\alpha,\theta)
 \end{equation}
\noindent
The Variance Gamma distribution is infinitely divisible. The Fourier transform function has an explicit closed-form in (\ref {eq:l5}).
\begin{equation}
 F[f](x) =\frac{e^{-i\mu x}}{\left(1+\frac{1}{2}\theta \sigma^{2}x^{2} + i\delta\theta x\right)^{\alpha}} \quad 
 f(y) =\frac {1} {\sigma\Gamma(\alpha) \theta^{\alpha}}\int_{0}^{+\infty} \frac{1}{\sqrt{2\pi v}}e^{-\frac{(y-\mu-\delta v)^2}{2v\sigma^2}}v^{\alpha -1}e^{-\frac{v}{\theta}} \,dv \label {eq:l5} 
  \end{equation}
When $\delta=0$, we have Symmetric Variance Gamma (SVG) Model. It can be shown by Cumulant-generating function\cite{kendall1946advanced} that
\begin{equation}
E(X)=\mu \quad 
Var(X)=\alpha \theta \sigma^{2} \quad 
Skew(X)=0 \quad  Kurt(X)=3(1+\frac{1}{\alpha}) \label{eq:l6}
 \end{equation} 
Fig \ref{fig21}, Fig \ref{fig22} and Fig \ref{fig23} display the FRFT estimations of the probability density function with Parameter values: $\mu=-2$, $\delta=0$, $\sigma=1$, $\alpha=1$, $\theta=1$. As shown in Fig \ref{fig22}, the probability density is left asymmetric and right asymmetric when the parameter ($\delta$) is negative and positive respectively. For $\delta=0$, the density function is symmetric, as shown in $(\ref {eq:l6})$. The shape parameter ($\alpha$) impacts the peakedness and tails of the distribution, as illustrated in Fig \ref{fig23} and $(\ref {eq:l6})$; heavier is the tails, shorter is the peakedness. $\theta$ and $\sigma$ have the same impact on the distribution. As shown in $(\ref {eq:l6})$, both change only the variance.
 
\section{Variance Gamma (VG) Model  }
\subsection{Model for asset Price}
\noindent
The VG model was introduced by Madan \cite{madan1990variance}. The asset price is modeled on business time $(k)$ as follows. $\mu, \delta \in R$, $\sigma>0$, $\alpha>0$ and $\theta>0$
 \begin{equation}
 Y_{k}=\mu + \delta V_{k} +\sigma \sqrt{V_{k}}Z  \quad  \quad Z \sim N(0,1)  \quad V_{k}\sim \Gamma(\alpha,\theta) 
\label{eq:l81}
 \end{equation}
 \begin{equation}
S_{K}=S_{k-1}e^{\sum_{j=k}^{K}Y_{j}}   \quad  \quad    T_{K}=\sum_{k=1}^{K}V_{k} 
\label{eq:l82}
 \end{equation}
\noindent 
$\{T_{k}\}$ is the activity time process, a non-negative stationary independent increment, called the subordinator.  $\mu$ is the drift of the physical time scale $t$, $\delta$ is the drift of the activity time process, and $\sigma$ is the volatility. The density of $Y_{j}$ and its Fourier transform were provided in (\ref {eq:l5}). See \cite{nzokem2021fitting}, Appendix A.1, for proof of (\ref {eq:l5}).\\
$Y_{k} $ is the return variable of the stock or index  price, we have (\ref{eq:l21a}) from (\ref{eq:l81}) and (\ref{eq:l82}).
\begin{equation}
Y_{k} =log(\frac{S_{k}}{S_{k-1}})  \quad  \quad  E(Y_{k}|V_{k}) \sim N(\mu +\delta V_{k}, \sigma \sqrt{V_{k}}) \quad \quad V_{k}\sim \Gamma(\alpha,\theta) \label{eq:l21a} 
 \end{equation} 
For $\alpha=\frac{1}{\theta}$  and $0<\theta<<1$, $Y_{k}$ in (\ref{eq:l81}) becomes (\ref{eq:l81a}).    \begin{equation}
Y_{k}=\mu + \sigma Z  \quad  \quad Z \sim N(0,1)  \label {eq:l81a}
 \end{equation}
 The Classical Lognormal Model (CLM) in (\ref {eq:l81a}) is a special case of the Variance - Gamma Model. See \cite{nzokem2021fitting}, Appendix A.1, for proof of (\ref {eq:l81a}).\\
 
\subsection{SPY ETF data}

\begin{figure}[ht]
\vspace{-0.4cm}
\centering
	 \begin{minipage}[b]{0.4\linewidth}
	    \vspace{-0.2cm}
	   \centering
	   \includegraphics[width=\textwidth]{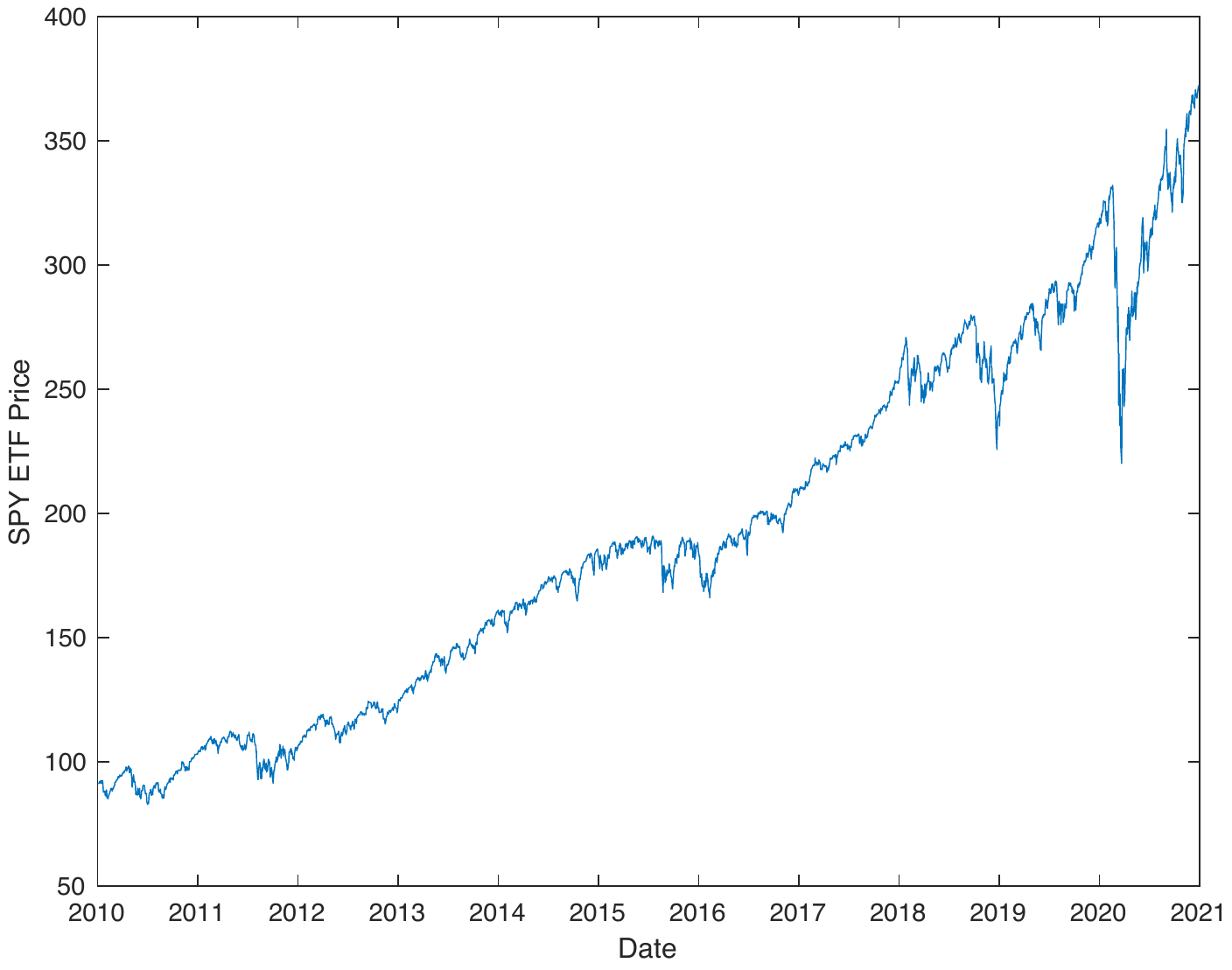}
	    \vspace{-0.7cm}     
    \caption{Daily SPY ETF Price }
         \label{fig31}
	\end{minipage}
\hspace{-0.3cm} 
\centering
	 \begin{minipage}[b]{0.4\linewidth}
	 \vspace{-0.2cm}
	   \centering
	   \includegraphics[width=\textwidth]{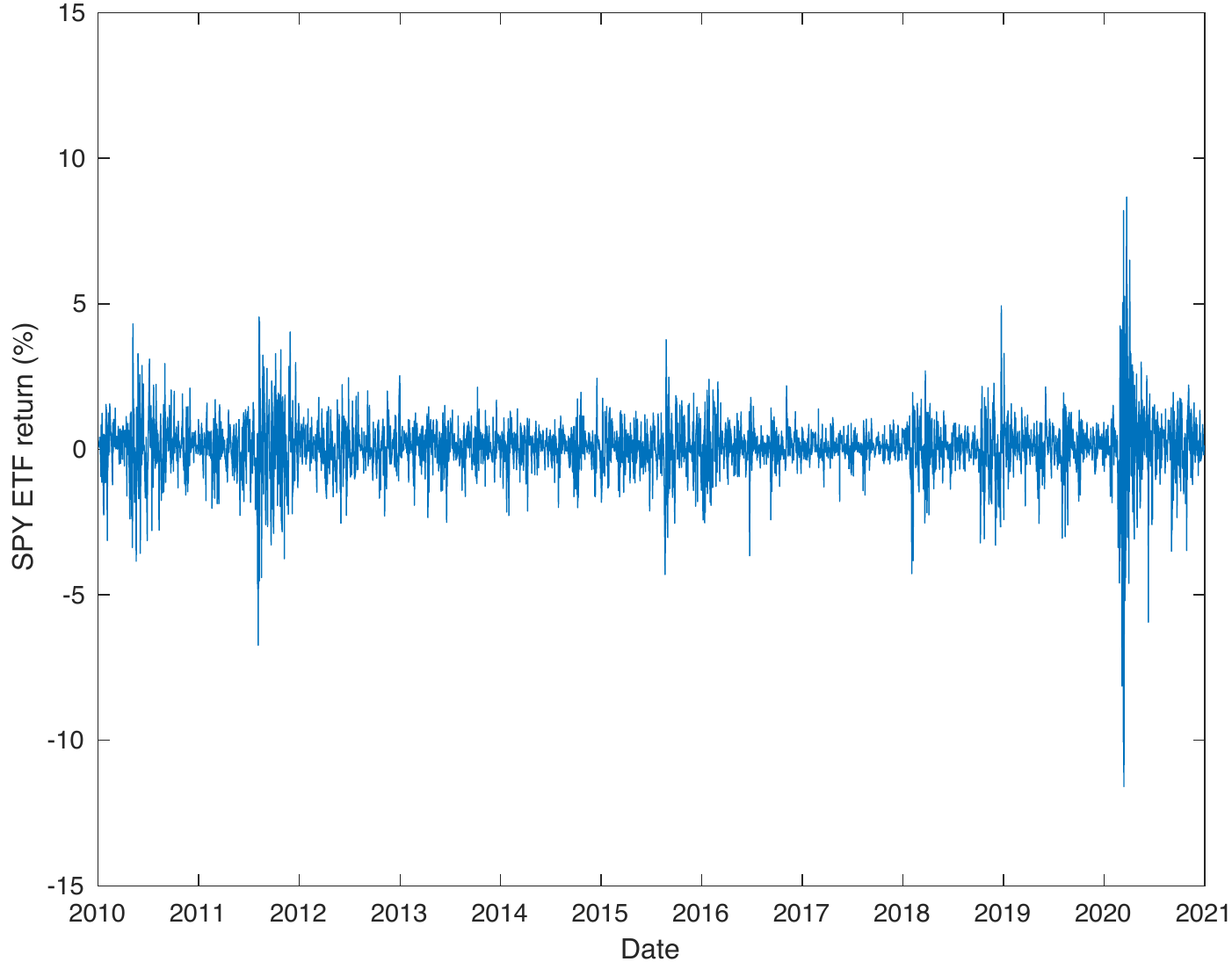}
	    \vspace{-0.7cm}
    \caption{Daily SPY ETF return }
         \label{fig32}
	\end{minipage}
	\vspace{-0.3cm}
\end{figure}

\noindent 
The data comes from the SPY ETF, called SPDR S\&P 500 ETF (SPY). The SPY is an Exchange-Traded Fund (ETF) managed by State Street Global Advisors that tracks the Standard \& Poor's 500 index (S\&P 500 ), which comprises 500 large and mid-cap US stocks. The SPY ETF is a well-diversified basket of assets listed on the New York Stock Exchange (NYSE). Like other ETFs, SPY ETF provides the diversification of a mutual fund and the flexibility of a stock.\\
\noindent 
The SPY ETF data was extracted from Yahoo finance. The daily data was adjusted for splits and dividends. The period spans from January 4, 2010, to December 30, 2020. 2768 daily SPY ETF prices were collected, around 252 observations per year, over 10 years. The dynamic of daily adjusted SPY ETF price is provided in  Fig \ref{fig31}.\\
 \noindent
 Let the number of observations $N=2768$, and the daily observed SPY ETF price $S_{j}$ on day $t_{j}$ with $j=1,\dots,N$; $t_{1}$ is the first observation date (January 4, 2010) and $t_{N}$ is the last observation date (December 30, 2020). The daily SPY ETF log return $(y_{j})$ is computed as in (\ref{eq:l8}).\\
\begin{equation}
y_{j}=\log(S_{j}/S_{j-1}) \hspace{10 mm}  \hbox{ $j=2,\dots,N$}\label{eq:l8}\\
    \vspace{0.2cm}
 \end{equation}
\noindent  
The results of the daily SPY ETF return  are shown in Fig \ref{fig32}.\\

\noindent
As shown in Fig \ref{fig31} and Fig \ref{fig32}, like other stocks and securities, SPY ETF was unusually volatile in the first quarter of $2020$  amid the coronavirus pandemic and massive disruptions in the global economy. $13$ daily return observations were identified as outliers and removed from the data set in order to avoid a negative impact on the statistics and estimators.
\section{Variance Gamma (VG) Model Estimations}
From a probability density function $f(y,V)$ with parameter $V$ of size ($p=5$) and the sample data $Y$ of size ($M=2755$),  we define the Likelihood function and its derivatives.
 \begin{equation}
l(y,V) = \sum_{j=1}^{M} log(f(y_{j},V))
\label{eq:l91}
\vspace{-0.3cm}
 \end{equation}
 \begin{equation}
\frac{dl(y,V)}{dV_j} = \sum_{i=1}^{M} \frac{\frac{df(y_{i},V)}{dV_j}}{f(y_{i},V)}
\label{eq:l92}
 \end{equation}
 \begin{equation}
  \frac{d^{2}l(y,V)}{dV_{k}dV_{j}} = \sum_{i=1}^{M} \left(\frac{\frac{d^{2}f(y_{i},V)}{dV_{k}dV_{j}}}{f(y_{i},V)}- \frac{\frac{df(y_{i},V)}{dV_{k}}}{f(y_{i},V)}\frac{\frac{df(y_{i},V)}{dV_j}}{f(y_{i},V)}\right)
  \label{eq:l93}
 \end{equation}
\noindent
With $1\leq k\leq p \ \ and \ \ 1\leq j \leq p$.\\
 $f(y_{i},V)$, $\frac{df(y_{i},V)}{dV_j}$, $\frac{d^{2}f(y_{i},V)}{dV_{k}dV_{j}} $ are computed with the FRFT on each $y_{i}$  with $1\leq i \leq M$. See Fig $4$, Fig $5$, Appendix B.2 and Appendix C.3  in \cite{nzokem2021fitting}, these figures display the shape of the quantities $\frac{df(y_{i},V)}{dV_j}$, $\frac{d^{2}f(y_{i},V)}{dV_{k}dV_{j}} $, which can be Odd or Even functions.\\
 \noindent
 The Newton Raphson Iteration process in (\ref{eq:l94}) was implemented on the score function ($I'(y,V)$), and the Fisher information matrix  ($I''(y,V)$).
     \begin{equation}
V^{n+1}=V^{n}+{\left(I''(y,V^{n})\right)^{-1}}I'(y,V^{n})\label {eq:l94}
 \end{equation}
\noindent
With initial value $\sigma=\alpha=\theta=1$, $\delta=\mu=0$, the maximization procedure convergences after  21 iterations  for Asymmetric Variance-Gamma Model  (AVG). The result of the iteration Process (\ref{eq:l94})  are shown in Table \ref{tab1}.\\
 \begin{table}[ht]
\vspace{-0.9cm}
\caption{Results of AVG Model Parameters Estimations}
 \label{tab1}
\centering
\resizebox{13cm}{!}{%
\begin{tabular}{cccccccc}
\br
\textbf{Iterations} &
  \textbf{$\mu$} &
  \textbf{$\delta$} &
  \textbf{$\sigma$} &
  \textbf{$\alpha$} &
  \textbf{$\theta$} &
  \textbf{$l(y,V)$} &
  \textbf{$||\frac{dl(y,V)}{dV}||$} \\
  \br
1  & 0          & 0          & 1          & 1          & 1          & -3582.8388 & 598.743231 \\
2  & 0.05905599 & -0.0009445 & 1.03195903 & 0.9130208  & 1.03208412 & -3561.5099 & 833.530396 \\
3  & 0.06949925 & 0.00400035 & 1.04101444 & 0.88478895 & 1.05131996 & -3559.5656 & 447.807305 \\
4  & 0.07514039 & 0.00055771 & 1.17577397 & 0.67326429 & 1.17778666 & -3569.6221 & 211.365781 \\
5  & 0.08928373 & -0.0263716 & 1.03756321 & 0.83842661 & 0.94304967 & -3554.4434 & 498.289445 \\
6  & 0.08676498 & -0.0521887 & 1.03337015 & 0.85591875 & 0.95066351 & -3550.6419 & 204.467192 \\
7  & 0.086995   & -0.0608517 & 1.02788937 & 0.87382621 & 0.95054954 & -3549.8465 & 66.8039738 \\
8  & 0.08542912 & -0.058547  & 1.02705241 & 0.88258411 & 0.94321299 & -3549.7023 & 15.3209117 \\
9  & 0.08478622 & -0.0576654 & 1.02995166 & 0.88447791 & 0.93670036 & -3549.6921 & 1.14764198 \\
10 & 0.08477798 & -0.0577736 & 1.02922308 & 0.88449072 & 0.93831041 & -3549.692  & 0.17287708 \\
11 & 0.08476475 & -0.0577271 & 1.02960343 & 0.88450434 & 0.93755549 & -3549.692  & 0.07850459 \\
12 & 0.08477094 & -0.0577488 & 1.02942608 & 0.8844984  & 0.93790784 & -3549.692  & 0.03723941 \\
13 & 0.08476804 & -0.0577386 & 1.02950937 & 0.88450117 & 0.93774266 & -3549.692  & 0.01732146 \\
14 & 0.0847694  & -0.0577434 & 1.02947043 & 0.88449987 & 0.93781995 & -3549.692  & 0.00813465 \\
15 & 0.08476876 & -0.0577411 & 1.02948868 & 0.88450048 & 0.93778375 & -3549.692  & 0.00380345 \\
16 & 0.08476906 & -0.0577422 & 1.02948014 & 0.88450019 & 0.9378007  & -3549.692  & 0.00178206 \\
17 & 0.08476892 & -0.0577417 & 1.02948414 & 0.88450033 & 0.93779276 & -3549.692  & 0.00083415 \\
18 & 0.08476898 & -0.0577419 & 1.02948226 & 0.88450026 & 0.93779648 & -3549.692  & 0.00039063 \\
19 & 0.08476895 & -0.0577418 & 1.02948314 & 0.88450029 & 0.93779474 & -3549.692  & 0.00018289 \\
20 & 0.08476897 & -0.0577419 & 1.02948273 & 0.88450028 & 0.93779555 & -3549.692  & 8.56E-05   \\
21 & 0.08476896 & -0.0577418 & 1.02948292 & 0.88450029 & 0.93779517 & -3549.692  & 4.01E-05  \\
\br
\end{tabular}
}
\vspace{-0.3cm}
\end{table}

\noindent
The estimation of other models are summarized in Table \ref{tab2}. The method of moments provides the initial values for AVG1 and SVG1 maximization procedure. The results are labeled AVG1 for Asymmetric VG Model and SVG1 for Symmetric VG Model. Another initial value was chosen: $\sigma=\alpha=\theta=1$, $\delta=\mu=0$. The results are labeled AVG2 and SVG2 respectively for Asymmetric VG and Symmetric VG Models.\\The Maximum Likelihood estimations  are summarized in Table \ref{tab2}.\\
\begin{table}[ht]
\vspace{-1cm}
\caption{Variance-Gamma  Parameters Estimations}
\label{tab2} 
\vspace{-0.3cm}
\begin{center}
\begin{tabular}{cccccc}
\br
\textbf{Model} & \textbf{$\mu$} &\textbf{$\delta$} &  \textbf{$\sigma$} &  \textbf{$\alpha$} &   \textbf{$\theta$} \\
\br
\textbf{AVG1}&{$0.1683$}&{$-0.1089$}&{$0.8987$}&{$0.9284$}&{$1.0546$}\\
\textbf{SVG1}&{$0.0510$}&{$ $}&{$0.9378$}&{$0.8490$}&{$1.0929$}\\
\textbf{AVG2}&{$0.0848$}&{$-0.0577$}&{$1.0295$}&{$0.8845$}&{$0.9378$}\\
\textbf{SVG2}&{$0.0652$}&{$ $}&{$0.9939$}&{$0.8770$}&{$0.9937$}\\
\textbf{CLM}&{$0.0541$}&{$ $}&{$0.9740$}&{$ $}&{$ $}\\
\br
\end{tabular}
\end{center}
\vspace{-0.8cm}
\end{table}

\noindent
 The estimation of parameters ($\mu$, $\sigma$) of the Classical Lognormal Model (CLM) was added to Table \ref{tab2}. 
 
\section{Comparison of Variance Gamma (VG) Models}
\begin{figure}[ht]
 \vspace{-0.7cm}
     \centering
         \includegraphics[scale=0.45] {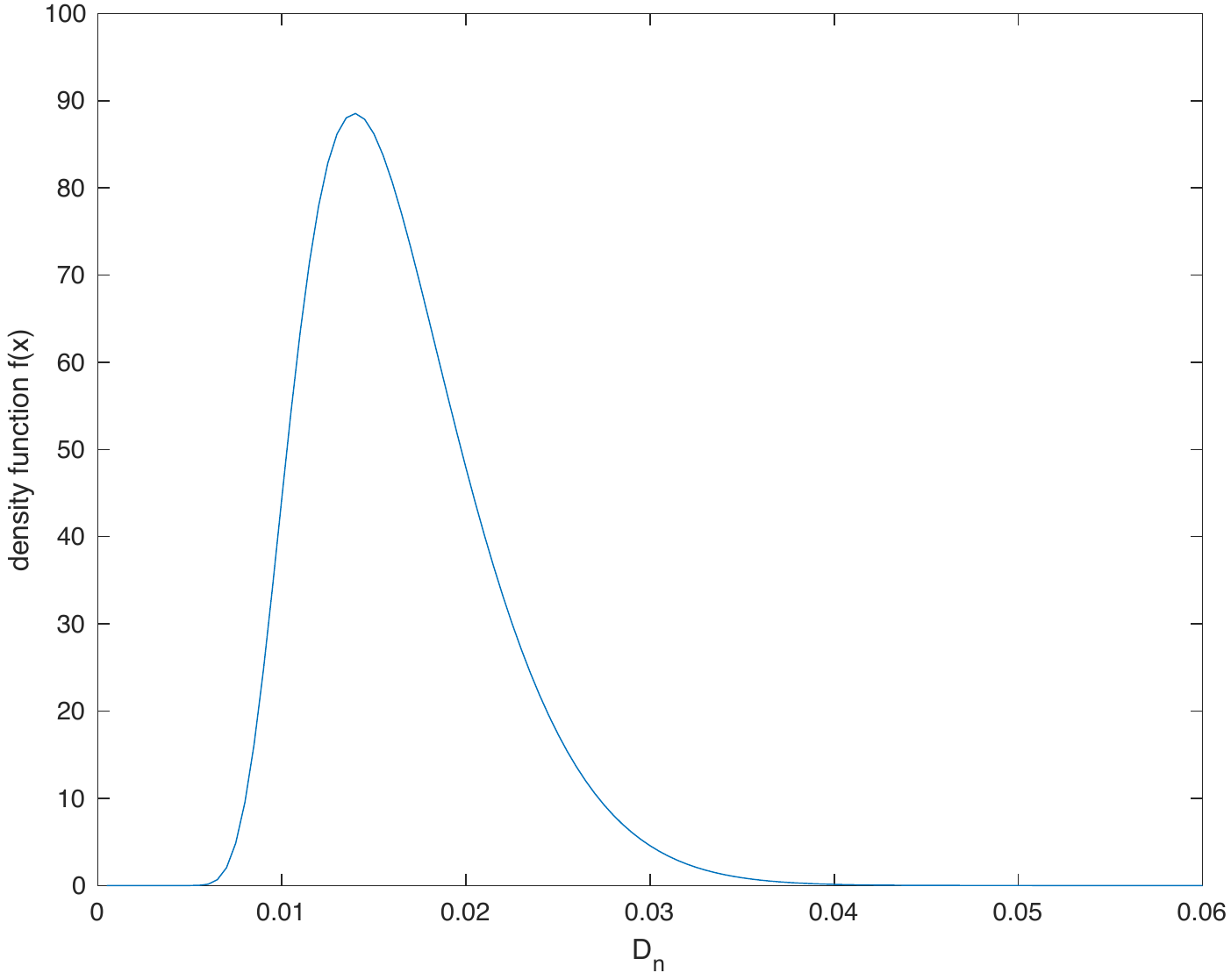}
 \vspace{-0.5cm}
        \caption{Kolmogorov-Smirnov Estimator ($\hat{D_{n}}$) probability density ($n=2755$)  under the null hypothesis $H{0}$}
        \label{fig5}
 \vspace{-0.7cm}
\end{figure}

\noindent
which VG model estimation fits the empirical distribution was also considered.  The Kolmogorov-Smirnov (KS) test was performed under the null hypothesis (H0) that the sample $\{y_{1}, y_{2}\dots y_{n}\}$ comes from VG model. The Kolmogorov-Smirnov (K-S) estimator ($D_{n}$) is defined in (\ref {eq:l10}). 
\begin{equation}
\vspace{-0.3cm}
D_{n} = \sup_{x}{|F(x)-F_{n}(x)|} 
\quad 
P_{value} = prob(D_{n}>d_{n} |H_{0})  \label{eq:l10} 
 \end{equation}
\noindent 
$F_{n}(x)$ denotes the empirical cumulative distribution and $n$ is the sample size. The VG cumulative distribution function ($F$) was computed with FRFT from its Fourier. \\
\noindent
The cumulative distribution of $D_{n}$\cite{dimitrova2020computing}  under the null hypothesis was computed and the density function was deduced. The computed density function is shown in Fig \ref{fig5}. Under the null hypothesis (H0), $D_{n}$ has a positively skewed distribution with means ($\mu=0.0165$) and standard deviation  ($\sigma=5*10^{-3}$).\\

\noindent
$d_{n}$ is the value of the KS estimator ($D_{n}$) computed from the sample $\{y_{1}, y_{2}\dots y_{n}\}$. Based on \cite{krysicki1999rachunek,kucharska2009nig}, $d_{n}$  can be estimated as  follows.
\begin{equation}
d^{+}_{n}= \sup_{0\leq j\leq P}{|F(x_{j})-F_{n}(x_{j})|}  \quad 
d^{-}_{n}= \sup_{1\leq j\leq P}{|F(x_{j})-F_{n}(x_{j-1})|} \quad
d_{n}= Max(d^{+}_{n}, d^{-}_{n}) \label{eq:l40}
 \end{equation} 
The statistics estimation for SVG2 model is shown in \cite{nzokem2021fitting}.  $d^{-}_{n}=max((1))=0.023629$, $d^{+}_{n}=max((2))=0.021986$ and $d_{n}=0.023629$. See Appendix E.5, Table E.5 in \cite{nzokem2021fitting}.\\

\noindent
For each model, KS-Statistics ($d_{n}$) and P\_values were computed, and the results are provided in Table \ref{tab2}.
\begin{table}[ht]
\vspace{-0.5cm}
\caption{\label{tab3} Kolmogorov-Smirnov (KS) test}
\vspace{-0.3cm}
\begin{center}
\begin{tabular}{lcl}
\br
\textbf{Model} & \textbf{KS-Statistics ($d_{n}$)} & \textbf{P\_values} \\
\br
\textbf{AVG1} & 0.054290 & 0.00001691\% \\
\textbf{SVG1} & 0.036763 &0.1136\% \\
\textbf{AVG2} & 0.028182 & 2.4668\%  \\
\textbf{SVG2} &0.023629 & 9.0788\%    \\
\textbf{CLM}  & 0.095791 & 0\%             \\
\br
\end{tabular}
\end{center}
\vspace{-0.8cm}
\end{table}

\begin{figure}[ht]
\vspace{-0.2cm}
	 \begin{minipage}[b]{0.32\linewidth}
	    \vspace{-0.2cm}
	   \centering
	   \includegraphics[width=\textwidth]{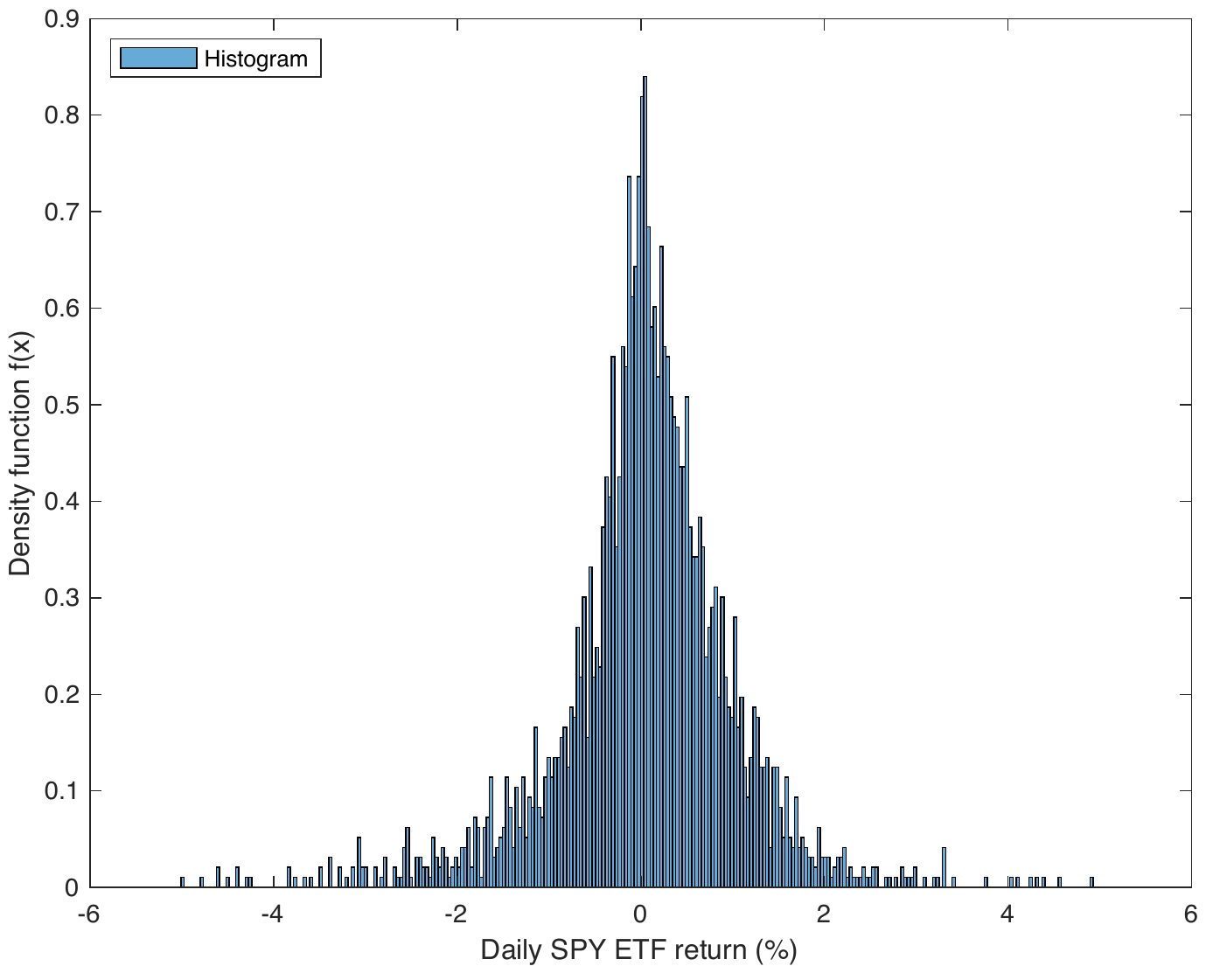}
	    \vspace{-0.9cm}     
    \caption{return Histogram}
         \label{fig41}
	\end{minipage}
 \hfill 	
	 \begin{minipage}[b]{0.32\linewidth}
	 \vspace{-0.2cm}
	   \centering
	   \includegraphics[width=\textwidth]{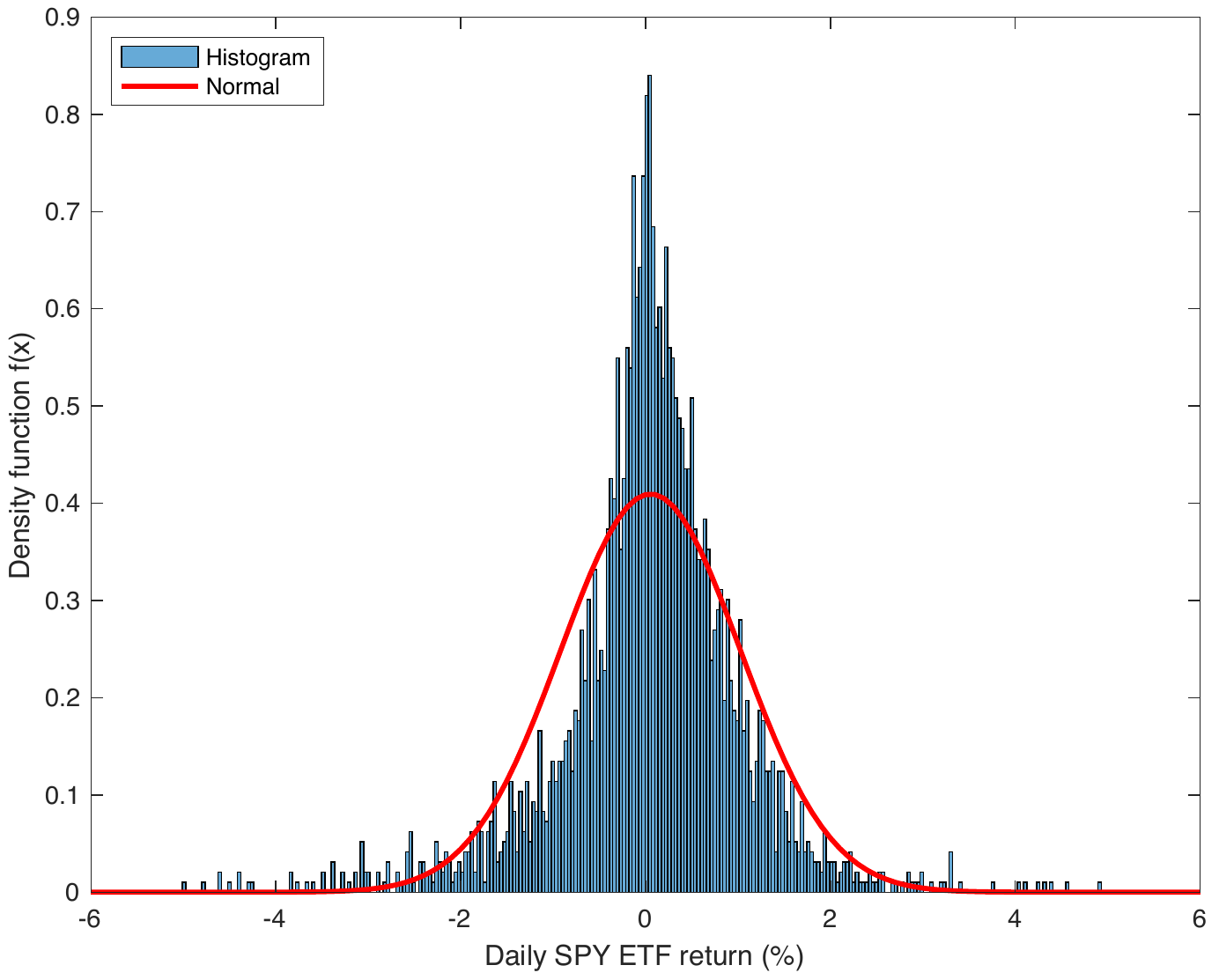}
	    \vspace{-0.9cm}
    \caption{CLM Model  }
         \label{fig42}
	\end{minipage}
 \hfill	
	 \begin{minipage}[b]{0.32\linewidth}
	 \vspace{-0.2cm}
	   \centering
	   \includegraphics[width=\textwidth]{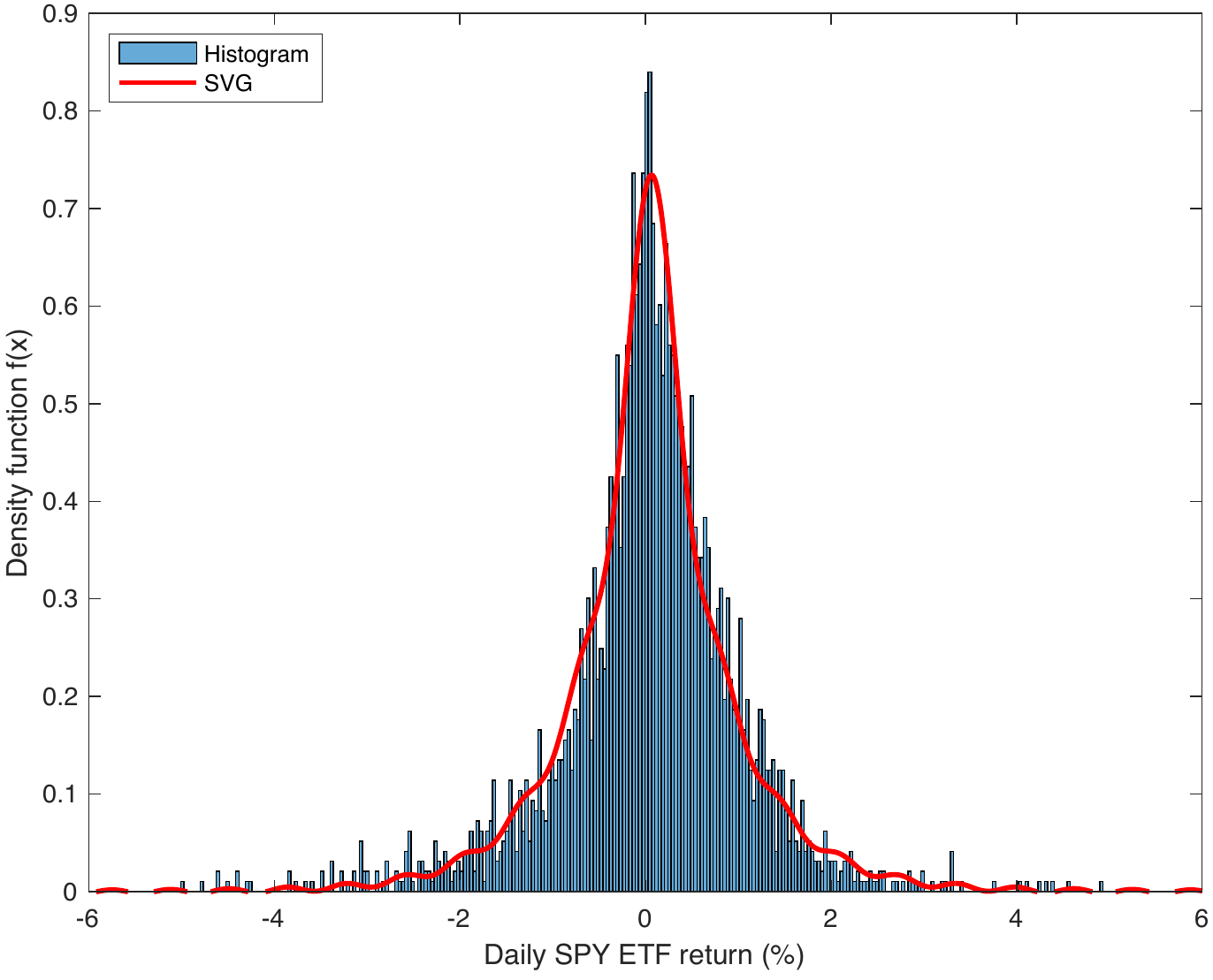}
	   \vspace{-0.9cm}
     \caption{ SVG2 Model}
         \label{fig43}
	\end{minipage}
\vspace{-0.5cm}
\end{figure}

\noindent
Most KS-statistic $d_{n}$ has high value and suggests that the sample $\{y_{1}, y_{2}\dots y_{n}\}$ rejects the null hypothesis (H0), except SVG2 model, and to some extent, AVG2 model at $2\%$ risk level.\\
\noindent
As shown in Table \ref{tab3}, VG models from the method of moments do not fit the sample data distribution. The $P\_values$ is less than $5\%$, and the null hypothesis $H_{0}$ can not be accepted at that risk level. The CLM does not fit the sample data distribution at $5\%$. Regarding the maximum likelihood method, the SVG2 model has $P\_values=9.079\%$, which is high than the classical threshold  $5\%$. Therefore, SVG2 model can not be rejected. See \cite{nzokem2021fitting} for $d_n$ and $P\_values$ computations.\\ 
\noindent  
The daily SPY ETF return histogram was compared to the density function of two models (SVG2, CLM) as shown in Fig \ref{fig42} and Fig \ref{fig43}. It results that the peakedness of the histogram explains the high level of the KS-Statistics in Table \ref{tab3} and the model rejection.\\
\noindent

 For work related to Normal and exponential distributions, see \cite{nzokem2021sis, aubain2020,Nzokem2020EpidemicDA,aubain2021}

\section {Conclusion} 
\noindent 
In the study, the FRFT-based technique is used to compute and analyze the probability density function of the Variance-Gamma (VG) model; and perform the estimation of the five parameters of the VG model. The results show that the VG model captures the peakedness and leptokurtosis properties of the daily SPY sample data. The findings provide evidence that the VG model fits better than the CLM Model. The Kolmogorov-Smirnov (KS) goodness-of-fit test shows that the Maximum Likelihood method with FRFT produces a good estimation of the VG model, which fits the empirical distribution of the sample data. 
\section*{References}
\bibliographystyle{iopart-num}
\bibliography{JPCS_VGM.bib}

\end{document}